\documentclass[final]{aipproc}
\layoutstyle{6x9}
\begin{document}
\title{Dyson-Schwinger Equations and Coulomb Gauge Yang-Mills Theory}
\author{P.~Watson}{address={Institut f\"ur Theoretische Physik, Auf der Morgenstelle 14, D-72076 T\"ubingen, Germany}}
\author{H.~Reinhardt}{address={Institut f\"ur Theoretische Physik, Auf der Morgenstelle 14, D-72076 T\"ubingen, Germany}}
\begin{abstract}
Coulomb gauge Yang-Mills theory is considered within the first order formalism.  It is shown that the action is invariant under both the standard BRS transform and an additional component.  The Ward-Takahashi identity arising from this non-standard transform is shown to be automatically satisfied by the equations of motion.
\end{abstract}
\maketitle
Dyson-Schwinger equations [DSEs] are a natural tool for studying the continuum properties of field theories.  Applied to Quantum Chromodynamics [QCD] in the Landau gauge, significant progress has been made in understanding the nature of confinement, the infrared behaviour of propagators and the light hadron spectrum with good agreement, where pertinent, with lattice studies (see for example the recent review \cite{Fischer:2006ub} and references therein).  As with any such calculation, the question naturally arises whether the relevant observables are properly gauge invariant and this leads to the desire for a DSE study of Coulomb gauge QCD.  In Coulomb gauge there are many different existing studies, though perhaps the most closely related are those based on the Hamiltonian approach \cite{Feuchter:2004mk}, however, a true DSE study has not yet been attempted.  Here we outline some of the basic formalism for such a study of Yang-Mills theory within the first order formalism and highlight an interesting feature -- that there exists more than one BRS-type symmetry but that the `extra' invariance is equivalent to the equations of motion.  A more complete description of this work can be found in \cite{me}.

Let us consider the following functional integral (working in Minkowski space):
\[
Z=\int{\cal D}\Phi\exp{\left\{\imath{\cal S}\right\}}=\int{\cal D}\Phi\exp{\left\{\imath\int d^4x\left[\frac{1}{2}E^2-\frac{1}{2}B^2\right]\right\}}
\]
where $\vec{E}^a=-\partial_t\vec{A}^a-\vec{D}^{ab}A_0^b$ and $\vec{B}^a$ are the chromo-electric and -magnetic components of the field strength tensor $F_{\mu\nu}$ ($A$ is the gauge field, $\vec{D}$ is the covariant derivative and $\Phi$ denotes the collection of all fields).  If we fix to Coulomb gauge ($\vec{\nabla}\!\cdot\!\vec{A}=0$) with a Lagrange-multiplier field $\lambda^a$ using the Faddeev-Popov technique then the action gets the additional term
\[
{\cal S}_{fp}=\int d^4x\left[-\lambda^a\vec{\nabla}\!\cdot\!\vec{A}^a-\bar{c}^a\vec{\nabla}\!\cdot\!\vec{D}^{ab}c^b\right].
\]
where the fields $\bar{c}$ and $c$ are the ghost fields.  The ghost term can also be written as a functional determinant $\mathrm{Det}(-\vec{\nabla}\!\cdot\!\vec{D})$ in $Z$.  The action as it is so far written, is invariant under the standard BRS transform \cite{Marciano:1978ee}.  To convert to the first order formalism \cite{Zwanziger:1998ez}, we introduce $\vec{\pi}$-fields (classically, momentum conjugate to the $\vec{A}$-fields) via the following identity
\[
\exp{\left\{\imath\int\frac{1}{2}E^2\right\}}=\int{\cal D}\pi\exp{\left\{\imath\int\left[-\frac{1}{2}\pi^2-\vec{\pi}^a\!\cdot\!\vec{E}^a\right]\right\}}
\]
and to maintain BRS invariance (given that under an infinitessimal local transform parameterised by $\theta_x^a=c_x^a\delta\lambda$, the variation of $\vec{E}^a$ is $\delta\vec{E}^a=f^{abc}\theta_x^b\vec{E}^c$) we require that
\[
\delta\pi^a=f^{abc}\theta_x^b\left[(1-\alpha)\vec{\pi}^c-\alpha\vec{E}^c\right],
\]
where $\alpha$ is some constant.  Since $\alpha$ is unconstrained we have {\it two} invariances.  To complete, we split the $\vec{\pi}$-field into transverse and longitudinal parts with
\[
\mathrm{const}=\int{\cal D}\phi{\cal D}\tau\exp{\left\{-\imath\int\tau^a\left(\vec{\nabla}\!\cdot\!\vec{\pi}^a+\nabla^2\phi^a\right)\right\}}
\]
and translate $\vec{\pi}\rightarrow\vec{\pi}-\vec{\nabla}\phi$ to finally give
\begin{eqnarray}
Z&=&\int{\cal D}\Phi\exp{\left\{\imath{\cal S}_B+\imath{\cal S}_{fp}+\imath{\cal S}_\pi\right\}},\nonumber\\
{\cal S}_\pi&=&\int\left[-\tau^a\vec{\nabla}\!\cdot\!\vec{\pi}^a-\frac{1}{2}\left(\vec{\pi}-\vec{\nabla}\phi\right)^2+\frac{1}{2}\left(\vec{\pi}^a-\vec{\nabla}\phi^a\right)\!\cdot\!\left(\partial_t\vec{A}^a+\vec{D}^{ab}A_0^b\right)\right].\nonumber
\end{eqnarray}
The (non-standard) $\alpha$-dependent part of the BRS transform reads ($\vec{X}^c=\vec{\pi}^c-\vec{\nabla}\phi^c-\partial_t\vec{A}^c-\vec{D}^{cd}A_0^d$)
\[
\delta\pi^a=f^{abc}\theta_x^b\vec{X}^c+\vec{\nabla}\delta\phi^c,\;\;\;\;\delta\phi^a=f^{abc}\frac{\vec{\nabla}}{(-\nabla^2)}\!\cdot\!\vec{X}^c\theta_x^b,
\]
with all other fields unaltered.

At this stage, it is pertinent to motivate the use of the first order formalism and this lies in the ability, at least formally, to reduce to transverse $\vec{A}$ and $\vec{\pi}$ degrees of freedom \cite{Zwanziger:1998ez}.  Those fields occuring linearly in the action may be integrated out, leaving functional $\delta$-functions, one of which implements Gau\ss' law:
\[
\delta(-\vec{\nabla}\!\cdot\!\vec{D}^{ab}\phi^b-gf^{ade}\vec{A}^d\!\cdot\!\vec{\pi}^e).
\]
Defining the inverse Faddeev-Popov operator $M$ via $-\vec{\nabla}\!\cdot\!\vec{D}^{ab}M^{bc}=\delta^{ac}$, the $\delta$-function now reads
\[
\mathrm{Det}^{-1}(-\vec{\nabla}\!\cdot\!\vec{D})\delta(\phi^a-M^{ac}gf^{cde}\vec{A}^d\!\cdot\!\vec{\pi}^e)
\]
and the inverse determinant {\it cancels} the Faddeev-Popov determinant exactly.  The energy divergences associated with the static ghost propagators in Coulomb gauge are automatically removed \cite{Zwanziger:1998ez}.  We are left with
\begin{eqnarray}
Z&=&\int{\cal D}\Phi\delta\left(\vec{\nabla}\!\cdot\!\vec{A}\right)\delta\left(\vec{\nabla}\!\cdot\!\vec{\pi}\right)\exp{\left\{\imath{\cal S}_B+\imath{\cal S}_\pi\right\}},\nonumber\\
{\cal S}_\pi&\sim&\int\left[-\frac{1}{2}\pi^2-\frac{1}{2}\left(\vec{A}\!\cdot\!\vec{\pi}\right)M(-\nabla^2)M\left(\vec{A}\!\cdot\!\vec{\pi}\right)+\vec{\pi}\!\cdot\!\partial_t\vec{A}\right]\nonumber
\end{eqnarray}
which expresses the theory in terms of transverse gluon degrees of freedom.  Since the action is now non-local these manipulations can be regarded as merely formal but do give insight into the previous, local formulation.

Each continuous transform of the theory may be regarded as a change of variables in the functional integral and that the action is invariant leads to a Ward-Takahashi identity.  In the case of the $\alpha$-dependent part of the BRS transform the Jacobian factor is trivial \cite{me} and we have that
\begin{equation}
\label{1}0=\int{\cal D}\Phi f^{abc}\vec{X}^c\!\cdot\!\left[\vec{K}^a-\frac{\vec{\nabla}}{(-\nabla^2)}\left(\kappa^a-\vec{\nabla}\!\cdot\!\vec{K}^a\right)\right]\exp{\left\{\imath{\cal S}\right\}}
\end{equation}
where $\vec{K}$ and $\kappa$ are sources introduced for the $\vec{\pi}$ and $\phi$ fields respectively.  In addition, for each field, from the observation that the integral of a total derivative vanishes (up to boundary terms which are here assumed to vanish) we have equations of motion -- the Dyson-Schwinger equations are functional derivatives of such equations.  For the $\vec{\pi}$ and $\phi$ fields respectively, these equations are:
\begin{eqnarray}
\label{2}\vec{K}^aZ\left[\vec{K},\kappa\right]&=&-\int{\cal D}\Phi\left[\vec{\nabla}\tau^a-\vec{X}^a\right]\exp{\left\{\imath{\cal S}\right\}}\\
\label{3}\kappa^aZ\left[\vec{K},\kappa\right]&=&\int{\cal D}\Phi\vec{\nabla}\!\cdot\!\vec{X}^a\exp{\left\{\imath{\cal S}\right\}}
\end{eqnarray}
where $Z\left[\vec{K},\kappa\right]$ is now the generating functional (the functional integral $Z$ from before in the presence of sources).  Inserting Eqs.~(\ref{2}) and (\ref{3}) into the right-hand side of Eq.~(\ref{1}) and using the antisymmetry property of the structure constant $f^{abc}$ it is immediately clear that the equation is automatically satisfied.  The two equations of motion and the $\alpha$-dependent part of the BRS transform give rise to exactly the same constraints on the functional integral and thus to any functional derivatives (i.e., Green's functions).

\begin{theacknowledgments}
Work supported by the DFG under contracts no. Re856/6-1 and Re856/6-2.
\end{theacknowledgments}
\bibliographystyle{aipproc}

\end{document}